%Paper: astro-ph/9211009
%From: tsvi@cfata3.harvard.edu (tsvi)
%Date: Fri, 13 Nov 92 11:23:55 -0500

entire .tex paper, including any non-standard tex or latex macros, and
instructions (at beginning) for stripping off any figures appended as
postscript files.
\magnification 1200
\baselineskip=20.pt
\def\etal{{\it etal.}}
\def\rf{\hfill\break\noindent}
\def\ppp{\par \smallskip \noindent \hangindent .5in \hangafter 1}

\def\ap{\approx}

\def\T{\tau}

\def\g{\gamma}
\def\a{\alpha}

\def\st{\sigma_T}
\def\pt{\ \ \ \ .}
\def\com{\ \ \ \ ,}

\def\G{\Gamma}

\bigskip
\hskip13 true cm
%version  Nov 8 accepted for publication in APJ Lett
\bigskip
\centerline{\bf FIREBALLS IN THE GALACTIC HALO AND $\gamma$-RAY BURSTS.}
\bigskip
\centerline{Tsvi Piran\footnote*{ Permanent Address: Racah Institute for
Physics, The Hebrew University, Jerusalem, Israel}}
\centerline{Harvard-Smithsonian Center for Astrophysics,}
\centerline{60 Garden Str. Cambridge, MA. 02138 USA.}
\centerline{and}
\centerline{Amotz Shemi}
\centerline{Dept. of Physics and Astronomy Tel Aviv
University,}
\centerline{Tel Aviv, Israel.}
\bigskip
\centerline{\bf Abstract.}

If gamma ray burst sources are in the galactic halo they inevitably
involve a creation of an opaque pair plasma fireball, just like in
cosmological sources.  We find that the typical physical conditions in a
galactic halo fireball are: optical depth $\ap 10^8$, thermal energy
$\ap 100 KeV$, maximal relativstic expansion  $\G \ap 300 $ and a
maximal baryonic load of $\ap 10^{-15} M_\odot$.  This does not rule
out galctic halo models but it poses an addtional severe constraint on
all such sources.  A comparison of these conditions with the physical
conditions at cosmological fireballs reveal that galactic halo
fireballs are less favorable than cosmological ones as sources of
gamma ray bursts.

\rf
\centerline{Accepted for Publication in Ap. J. Lett, Nov, 11, 1992.}
\rf
\centerline{{\it Subject Headings}:Gamma~Rays:~Bursts }
\vskip\eject

\bigskip
\centerline{\bf 1. Introduction}
\bigskip

The recent observation of the BATSE experiment on the COMPTON-GRO
observatory (Meegan \etal 1992) have demonstrated that gamma ray
bursts (GRBs) are distributed isotropically. The $V/V_{max}$ test
shows that the sources are not distributed homogenously. The
combination of both observations shows that
these sources cannot be local in the disk.  These
observations have led to two possible populations: cosmological
population (Usov \& Chibisov, 1975;
van den Bergh, S.  1983; Pac\'zynski, 1986, Piran, 1992;
Dermer, 1992; Mao \& Pac\'zynski, 1992) and a galactic halo population
(Katz, 1992; Lingenfelter \& Higdon 1992). All cosmological sources
have a common feature: they require a release of $\approx 10^{51}$ergs
in a short period of time.  A rise time of 10 msec suggests a source
dimension of less than 100 Km. Such sources will be extremely
optically thick due to pair plasma (the initial optical depth is
$\approx 10^{15}$) and they must pass a thermal phase. This fact was
used to criticize the cosmological models and it was argued that such
a system will cool and emits its energy in the x-ray range.  However,
the radiation and the pairs will behave like a fluid, forming a
relativistically expanding fireball (Goodman, 1986, Paczynski, 1986,
Shemi and Piran, 1990) and the nature of the escaping radiation will
be determined by various relativistic processes at the stage that the
fireball becomes optically thin or latter by its interaction with the
surrounding material (M\'es\'zaros and Rees 1992)

There have been several attempts to consider an alternative population
of galactic halo sources (Katz, 1992; Lingenfelter \& Higdon 1992;
Eichler and Silk, 1992; Li and Dermer, 1992).  GRBs sources in the
galactic halo must have a typical distance of $\ge 50$Kpc, to
accommodate the $V/V_{max}$ test and the remarkable isotropy in their
angular distribution.  An energy of $\ap 10^{41}$ergs is implied from
the observed fluencies of $\ap 10^{-7}{\rm ergs/cm}^2$.  It was hoped
that such sources will not have the optical thickness problem
associated with cosmological sources and that optically thin models,
constructed for much weaker sources at distances of several hundred pc
could also function in halo sources.  We show here that just like
cosmological GRBs, galactic halo GRBs must pass an optically thick
fireball phase.  We consider here two extreme cases of energy release:
an instantaneous release of $10^{41}$ergs into a volume of radius $R
\approx 10$Km and a steady state source (on time interval of $\ap$sec)
of a constant luminosity $L \ap 10^{41}$erg/sec.  We expect that
realistic sources are somewhere between these two extreme cases. In
both cases prolific pair production, via $\g\g \rightarrow e^+e^-$,
inevitably leads to creation of
an opaque media, which we call a fireball.  We examine the conditions
in galactic halo fireballs and show that they are less favorable than
the conditions in cosmic fireballs for producing the observed
gamma-ray  bursts.

\bigskip
\centerline{\bf 2. Instantaneous Energy Release}
\bigskip

We first consider a fireball with a total radiation energy E and a
radius R that is in a thermal equilibrium. The temperature will be:
$$
T = 113 E_{41}^{1/4} R_6^{-3/4} \ {\rm KeV}
\eqno(1)
$$
where $E_{41}$ is in units of $10^{41}$ergs and $R_6$ is in units of
$10^6$ cm.  At this temperature there is a significant number of
$e^+-e^-$ pairs.  The pair number density in the $T  < m_e c^2$ range
is:
$$
n_{\pm} \approx { \sqrt{2}\ \over \pi^{3/2} } \left ({\hbar \over
m_{e}c}\right)^{-3}
\left ({T\over m_e c^2}\right )^{3\over 2}\exp{\left [-{m_e c^2 \over T}
\right ]} = 3.8  \times 10^{29}
{T_{100}}^{3\over 2}\exp{\left[-{ 5.1 \over T_{100}}\right ]}
\ {\rm cm}^{-3} \ ,
\eqno (2)
$$
where $T_{100} = T/100$KeV.
The optical depth due to Compton scattering from these pairs is:
$$
a\tau_p = \st n_{\pm} R =
2.5 \times 10^{11}
{T_{100}}^{3\over 2}\exp{(-{5.1 \over T_{100}})} R_6 =
$$
$$
= 2.1 \times 10^{11} E_{41}^{3/8} R_6^{-1/8}
\exp{(-{ 4.5 R_6^{3/4} \over E_{41}^{1/4}})}  \pt
\eqno(3)
$$
For $E=10^{41}$ergs and $R= 10^6$cm the initial optical depth is
$\T_{p,0} = 2.2 \times 10^{9}$. For cosmological fireball with
$E=10^{51}$ergs the initial optical depth is even higher $ \T_{p,0} =
10^{15}$ The optical depth for pairs for Compton scattering is so
large (for both galactic halo and cosmic fireballs) that regardless of
the initial energy injection mechanism the fireball must become
thermal. The optical depth for $\g\g \rightarrow e^+e^-$ is of the
same order of magnitude or larger so even if pair do not exist
initially they will copiously form in a very short time.  Thus in
either case the initial phase of the gamma ray burst must be a thermal
fireball, characterized by an initial temperature $T_0$ given by Eq.
(1).

Goodman (1986) has shown that a radiation-pair fireball will expand
adiabatically as a fluid.  The expansion can be very well  understood
in terms of the behavior of a homogeneous radiation fireball (Shemi
\& Piran, 1990).  During the  expansion  the random thermal energy is
converted into a bulk kinetic energy of the outflow.  The  fireball
cools with: $T = T_0  ({R_0 /  R} )$,
and the relativistic Lorentz factor, $\G$ of the bulk motion is:
$\G = {T_0/T} = {R/R_0}$.
As the temperature goes down  pair annihilation is no longer reversible
and the pair population decays  exponentially.  This phase ends
when the systems becomes optically thin,  at $T_{esc}$. The optically
thin radiation no longer behaves like a fluid and the photons escape
freely.  The escape temperature, $T_{esc,p}$ depends very weakly on
the total energy and the initial radius of the fireball.  For
fireballs in the energy range of $10^{41}-10^{51} ergs$, $\tau_p = 1$
at $T_{esc,p}\ap 15-21$KeV.

Although the comoving temperature is lower then the initial
temperature, an observer at infinity will see a blueshifted spectrum
due to the large relativistic bulk motion of the fireball.  For a pure
radiation fireball the typical energy of the escaping radiation will be of
the order of the {\it initial} temperature:
$$
T_{obs} = \G T_{esc,pairs} = T_0
\eqno(4)
$$
The initial temperature, $T_0$, of a cosmic fireball is of the order
of $10$MeV, while the initial temperature, $T_0$, of a galactic halo
fireball is of the order of $100$KeV or less.  The escape temperature
is essentially the same in both cases.  It follows that the
acceleration phase in the comic fireball is significantly longer and
while the asymptotic Lorentz factor $\G$ is $\approx 10^{2}-10^{4}$
for a cosmic fireball it is $\approx 10$ or less for a pure radiation
galactic halo fireball.

Contamination of the fireball by baryons affects the fireball in three
different ways. The most sensitive effect is the influence of the
additional electrons on the opacity. The  opacity due to electron
associated with the baryons, $\tau_g$, is:
$$
\tau_g = \st \Bigl({M \over m_p} \Bigr) \Bigl( {3 \over 4 \pi R^2}
\Bigr) = 2\times 10^{20} m R_6 ^{-2} \com
\eqno(5)
$$
where $M$  is the baryonic mass and $m \equiv M/M_\odot$.  The gas
opacity decreases as $ R^{-2}$, slower than the pair opacity, which
falls exponentially once the temperature is below 500KeV.  The escape
temperature will be  the minimum of $T_{esc,p}$ and $ T_{esc,g}$ which
is determined by the condition $\tau_g (R) = 1$:
$$ {T_{esc,g} \over
T_0} =  \left( {4 \pi \over 3}\right )^{1/2} \left ({\st \over m_p}
\right )^{-1/2} M^{-1/2} R_0 =
$$
$$
7.1 \times 10^{-11} m^{-1/2} R_{0,6}
= 3 \times 10^{-4} \eta^{1/2}
E_{41}^{-1/2} R_{0,6} \com \eqno(6)
$$
where $ \eta \equiv E/M c^2 = 5.6 \times 10^{-14} E_{41}/m$ and
$R_{0,6}$ is the initial radius in units of $10^6$cm.  For $ 2.9
\times 10^{-19} E_{41}^{1/2} R_{0,6}^{1/2}$ $T_{esc,g} < T_{esc,p}$
and the baryonic contamination determines the escape temperature and
the escape radius.  The acceleration phase of the fireball is longer
than for a free fireball and hence the final Lorentz factor $\G_f $,
which is $\approx T_0/T_{esc}$, increases like $m^{1/2}$ for very
small baryonic masses. At these levels of baryonic contamination
the energy remains in the photons and only a small fraction of
it is converted to a kinetic energy of the baryons.

If $m > 1.7 \times 10^{-16} E_{41}$ the baryons will also influence
the dynamics of the fireball.  For a homogeneous expanding fireball
we have an average Lorentz factor $\G (T)$ (Shemi and Piran, 1990):
$$
\G (T)  = {E_0 + Mc^2 \over E(T) + Mc^2} =
{\eta + 1 \over \eta (T / T_0) + 1} \com
\eqno(7)
$$
and $\G_f$ is reached at $T=T_{esc}$.

Using Eq. 6 for $T_{esc}/T_0$ yields that the
maximal $\G_f \approx 330 $ is reached
when $m \approx 1.7 \times 10^{-16} E_{41}$, at the transition region
from one regime to the other. In this case $\G_f
\approx \eta$ and $T_{esc} \approx 30$eV. The relation $\G_f \approx
\eta $ holds when the mass increases and $\eta$ decreases.
When $m > 1.7 \times 10^{-16} E_{41}$, $\G_f$ is no longer
proportional to $T_0/T_{esc}$ and the final observed temperature,
which is $\G_f T_{esc}$, falls below $T_0$. Most of
the initial thermal energy
of the fireball is converted now to the kinetic energy of the baryons
$E_k$:
$$
{E_k \over E_0} = { M \G (T) \over E_0}   =
{1 + 1/ \eta \over \eta (T / T_0) + 1} \com
\eqno(8)
$$
The baryons reach relativistic velocities only in a narrow mass range
i.e. for $1.7 \times 10^{-16} E_{41} < m < 5.6 \times 10^{-14} E_{41}$.
For larger mass loads ($m > 5.6 \times 10^{-14}E_{41}$) $ \eta < 1 $
and the fireball never becomes relativistic.  Finally, if $M >
4.7\times 10^{-11} M_{\odot} (R_6 T_0)^3$, the gas pressure is larger
than the radiation pressure and the fireball evolution is quite
different. In particular the temperature is no longer proportional to
$1/R$.

\bigskip
\centerline {\bf 3. Steady State Scenario }
\bigskip

In 2. we have considered instantaneous release of energy and we have
seen that galactic halo sources are optically thick. We consider now a
steady state source where the energy is injected continuously by some
nonthermal mechanism.  The main question, again, is whether  it
possible to release the required luminosity without having an
optically thick flow which will change the spectrum.

The compactness parameter $l$ of such source is (Guilbert Fabian \&
Rees 1983):
$$
l = {L \over R} \times {\st \over m_e c^3}
  = 2.7\times 10^6 L_{41} \ R_6^{-1}
\pt \eqno(9)
$$
We denote by $f_\g$ the fraction of primary photons with energy larger
than $2 m_e c^2$. The optical depth, $\tau_{\g\g} $ is approximately
$f_\g l $ and $\exp(-f_\g l) $ is the probability for a high energy
photon, produced by the source, to escape without interacting and
producing an electron - positron pair.  The fraction $f_\g$ varies
from one source to another and is different for each specific source.
Guilbert (1983), for example, have shown that a source with an initial
photon power law spectrum: $ n(\nu) \propto \nu^{-\a} $ in the
interval $ \nu < \nu_M$ (with $ m_e c^2 \ll h \nu_M$) will be
optically thick to $\g \g \rightarrow e^+ e^-$ if: $ l > [\a /(2 -\a)]
(h \nu_M /m_e c^2)^{2- \a} $.  The compactness parameter given by Eq.
9 is so large that steady state sources are optically thick to $ \g \g
\rightarrow e^+ e^- $ and the radiation cannot escape freely.  Again
the pair-plasma and the radiation behave like a fluid and the escape
of the radiation from the source is best described in terms of a wind
of the type studied by Paczynski (1986,1990).  Just as  in the
previous case the wind expands and cools with $T \propto R^{-1}$ and
$\gamma \propto R$ until the local temperature drops and the
electron-positron annihilate and disappear. For a pure radiation wind
we find again that $T_{obs} =
\gamma T_{esc} \approx T_0 $.

If matter is ejected along in the wind the final observed temperature
will be lower, due to loss of energy to the inertia of the matter and
we have (Paczynski 1990):
$$
T_{obs} \approx 100   (M_s/M_\odot)^{1/6} L_{41}^{9/4} \dot m_{-14}^{-8/3}
{\rm KeV} \eqno(10)
$$
where
$M_s$ is the the mass of the central object and $\dot m_{-14}
\equiv 10^{-14} M_{\odot}$/sec is the injection rate of contaminating
mass.  Clearly $\dot m_{-14} < 1$ is needed if  $T_{obs} \ap 0.2 -
0.4$.

\bigskip
\centerline {\bf 4. Beaming and Relativistic Motion}
\bigskip

Our results also hold if the radiation is emitted only in a narrow
beam with an opening angle $\theta$.  Beamed emission will reduce the
overall energy required by a factor of $\theta^2 /2 $.  However,
locally it will not change the conditions inside the beam. The same
considerations that were used to determine the optical depth in the
spherically symmetric case, discussed earlier, are valid locally now
and such beams will be optically thick, and their initial spectrum
will be erased due to thermalization.

It seems that the only way to avoid an optically thick source is by
incorporating substantial relativistic motion of the source itself.
The optical depth
of a steady state source with a spectral index $\a$ moving
toward the observer with a relativistic factor $\g$ is (Krolik and
Pier 1991):
$$
\T_{\g \g} \ap 6.5\times 10^6 (2\g)^{- (2+\a )} R_6
A_{12}^{-2}F_{-7} D_{100}^2
\eqno(11)
$$
where $F = F_{-7} \times 10^{-7}~ {\rm
erg sec}^{-1}{\rm cm}^{-2}$ is the observed flux, $A = 10^{12} \times
A_{12}~ {\rm cm}^2$, is the area of the emitting region and $D = 100
\times D_{100}$kpc is the distance to the source.  A source for the
observed GRBs with a spectral index $\a$ can be optically thin if $\g
> 30$ for $ \a = 2$ or $\g > 100 $ for $\a = 1$.  In this case the
photons that are emitted in the local frame of the source in the few
KeV range are blue shifted to the $\gamma$-ray range in the observer
frame. It seems that for galactic halo sources the optical depth
problem can be resolved if the source (or a substantial part of it) is
moving with an extreme relativistic velocity.  However, this solution
raises several other problems such as: What is the accelerating
mechanism that accelerates the GRB sources to such a high relativistic
velocities (the largest velocity observed in bulk motion in
galactic object is $\approx c/3$ in SS433)? The only possible source
for such a high velocity is the relativistic expansion of the fireball
itself. What is the source of the
huge kinetic energy required for the bulk motion of such sources? and
finally it will require a much larger number of events to take account
for the observed one. It seems that the only possible source
for such a high velocity is the relativistic expansion of the fireball
of the type that we were discussing earlier.

\bigskip
\centerline {\bf 5. Discussion and Conclusions}
\bigskip

We have shown that $\gamma$ ray burst sources located at the galactic
halo must be optically thick and involve creation of an opaque
pair-plasma fireball.  The optical depth both Compton scattering from
thermalized plasma (Eq. 3) and for pair production by primary photons
(Eq. 9) are extremely high. The latter means that even if initially
there the emission in non thermal and if there are no pairs initially
the system will quickly reach a thermal equilibrium.  The difference
between numerical values in Eq. 3 and Eq. 9  represent the difference
between the two cases we have considered: $(\tau _p / l) \ap (c \Delta
t /R_6) \ap 10^4$.

The typical initial temperature, $T_0$, of the optically thick
fireball is $\approx 100$ KeV. This is also the observed temperature
$T_{obs} = \G_f T_{esc}$ of a clean fireball, where typical final
Lorentz factors $\G_f$ are less than 10.  Very small amounts of
baryonic contamination $2.9 \times 10^{-19} (E_{41} R_{0,6})^{1/2}
< m \ll 1.7 \times 10^{-16} E_{41}$ dominates the opacity at the
escape moment and hence extend the acceleration phase, without
overloading the inertia of the system. This leads* to a decrease of
$T_{esc}$ and to an increase of $\G_f$ with $T_{obs}$ remaining
unchanged.  $\G_f$ reaches a maximal value of around 330 for $ m
\approx 1.7 \times 10^{-16}$.  When $m$ increases further $\G_f$
decreases, with $\G_f \approx \eta$.  If $m > 1.1 \times 10^{-15}
(E_{41} R_0)^{2/3}$ (or if $(dm/dt) > 4 \times 10^{-14} M_\odot /sec$
in the steady state approximation) over $90 \% $ of the injected
energy is absorbed by the baryons.  For higher $m$ or $(dm/dt)$ values
the radiation energy is almost completely transferred to kinetic
energy of the accelerated particles, which are injected into the outer
space accompanied with by a modest or weak UV - X-ray signal. The
Lorentz $\Gamma$ factor is not as large as in cosmological bursts and
hence these baryons cannot produce the burst via interaction with
interstellar material, as suggested by M\'es\'zaros and Rees (1992)
for the cosmological scenario.

$T \approx 100 $ KeV is an upper limit to the fireball temperature,
once it thermalizes.  Thus it is very difficult to explain the
observed very high energy photons from a galactic halo fireball.
Since the typical Lorentz $\G$ factor are quite modest in a galactic
halo fireball we do not expect large relativistic effects in this
case. Note that for a cosmological fireball the typical initial temperature
is usually larger than 1 MeV and it is easy to account for the
photons energies, however it is still not clear how a non-thermal
spectrum can emerge.

To conclude we compare galactic halo fireballs with fireballs at
cosmological distances $ \ap Gpc$. The latter need $E \ge 10^{50}$ to
account for the observed $\gamma$-ray fluencies.  The initial
temperature $T_0$ is generally higher in a cosmological fireballs
while the escape temperature is of the same order. Hence the final
Lorentz $\Gamma$ factor, which is of the order $T_0/T_{esc}$ is much
higher and might reach $ \ap 10^2 -10^4$ in cosmological fireballs.
For the same reason, the mean observed temperature can easily exceed
MeV in cosmological fireballs while it is expected to be of the order
of $100$KeV or less in galactic ones.

We thank Ramesh Narayan for discussions and Jonathan Katz
for helpful comments.  This work was supported by NASA grant NAG 5-1904.

\bigskip
\rf
{\bf References}
\bigskip
\rf
Cavallo, G., \& Rees, M.J. 1978, MNRAS, {\bf 183}, 359.
\ppp
Dermer, C. D. 1992,  Phys. Rev. Letters, {\bf 68}, 1799.
\ppp
Eichler, D. and Silk, J., 1992, Science, {\bf 257}, 937.
\ppp
Goodman, J. 1986, ApJ (Letters), {\bf 308}, L47.
\ppp
Guilbert, P.W., 1983, in {\it Positron - Electron Pairs in
Astrophysics}, AIP
Con. 101,
eds: Burns, M.L., Harding, A.K. \& Ramaty, R., 405.
\ppp
Guilbert, P.W., Fabian, A.C. \& Rees, M.J., 1983, MNRAS, {\bf 205}, 593.
\ppp
Katz, J.I., Astrophy \& Space Sci., in press, 1992;
\ppp
Krolik,J.H. \& Pier, E.A., 1991, ApJ, {\bf 373}, 277.
\ppp
Li, H. and Dermer, C.D. 1992, Nature, {\bf 359},  514.
\ppp
Lingenfelter, R.E., \& Higdon, J.C., Nature, {\bf 356}, 132, 1992.
\ppp
Mao, S. \& Paczynski, B. 1992, ApJ (Letters), {\bf 388}, L45.
\ppp
Meegan, C.A., Fishman, G.J., Wilson, R.B., Paciesas, W.S., Pendleton, G.N.,
Horack, J.M., Brock, M.N. \& Kouveliotou, C., 1992, Nature, {\bf 355} 143.
\ppp
M\'es\'zaros, P. \& Rees, M.J., 1992, ApJ in press.
\ppp
Paczynski,B. 1986, ApJ (Letters), {\bf 308}, L51.
\ppp
Paczynski,B. 1990, ApJ, {\bf 363}, 218.
\ppp
Piran, T. 1992, ApJ (Letters), {\bf 389}, L45.
\ppp
Piran, T., Narayan, R. \& Shemi, A. 1992, Gamma-Ray Bursts, Huntsville Al,
1991, eds. W. S. Paciesas \& G. J. Fishman, AIP press, 149..
\ppp
Shemi, A. \& Piran, T., 1990, ApJ (Letters) {\bf 365}, L55.
\ppp
Usov V. V. \& Chibisov, G. V., 1975, Sov. Astr.- AJ. {\bf 119}, 115.
\ppp
van den Bergh, S.  1983, {\it Astrophys. Space Sci.}, {\bf 97}, 385.
\end